\documentclass{PoS}

\newcommand{\beq}{\begin{equation}}
\newcommand{\eeq}{\end{equation}}
\newcommand{\bea}{\begin{eqnarray}}
\newcommand{\eea}{\end{eqnarray}}

\usepackage{epsfig}
\input epsf

\title{IR-conformal gauge theories and composite Higgs}

\ShortTitle{IR-conformal gauge theories and composite Higgs}

\author{\speaker{E. T. Tomboulis}
\\
         Dept. of Physics and Astronomy, University of California, Los Angeles\\
        Los Angeles, CA 90095, USA\\
        E-mail: \email{tomboulis@physics.ucla.edu}}

\abstract{The existence of non-trivial IR and UV fixed points in gauge theories as a function of the number of fermion flavors and bare coupling is discussed in the light of recent work. It is pointed out that in fact only a small subset of potential IR-conformal gauge theories, i.e. theories whose IR behavior is determined by an IR fixed point, has so far been examined. Recent lattice computations of the spectrum in some cases where existence of an IR fixed point is reasonably assured, however, reveal a non-QCD-like spectrum with the lightest states being scalars. It is suggested that these naturally light composite scalars provide a natural setting for the construction of new composite Higgs models. A schematic outline of such a model is given. }

\FullConference{QCD-TNT-III-From quarks and gluons to hadronic matter: A bridge too far?,\\
		2-6 September, 2013\\
		European Centre for Theoretical Studies in Nuclear Physics and Related Areas (ECT*), Villazzano, Trento (Italy)}

\begin{document}

\section{Introduction}
The search for non-trivial IR or UV fixed points (FP) in gauge theories as a function of the  fermion flavor number and gauge group representation has been the focus of considerable effort over the last several years \cite{DelD1}. 
Apart from their intrinsic quantum field theory interest, such studies are motivated by their potential application to  physics beyond the Standard Model (BSM). One such proposal is that of walking TC, where the number of flavors is such as to place a system just outside and below the lower end of a conformal window. Many other possibilities, however, exist for BSM physics involving non-trivial IR FP's. 
In fact, as we will note in the following, so far, only a small subset of possible IR-conformal gauge theories, i.e. theories whose long distance behavior is governed by an IR FP, has been, even partially, explored. 

Here we will first review and discuss what is known and the many open issues concerning the phase diagram, existence of FT's and the spectrum of states as a function of the number of flavors and the gauge coupling. Lattice computations of the spectrum in cases where the existence of an IR FP is reasonably certain reveal (at any finite conformality deformation) a non-QCD-like spectrum of composites with the scalar states as the lowest states.  
We suggest that the appearance of these naturally light (massless) scalar states 
in the spectrum of IR-conformal theories allows construction of a wide class of composite Higgs 
models by the direct coupling of the electroweak and other gauge interactions to the IR-conformal theory. 
Such models, which do no rely on any walking TC mechanism or a composite Higgs as a NG boson due to the formation of some strong dynamics condensate, may evade some of the usual fine-tuning problems. The possibility, in particular, of obtaining IR FT's at (relatively) weak couplings by appropriate choice of $N_f$ and $N_c$ may be crucial for naturally obtaining  weakly-coupled SM Higgs and dark matter sectors.

\section{Phase diagram in $g$, $N_f$ and FP's}  
Recall that for a theory with $N_f$ fermion flavors in representation $R_f$ of a simple color group $G$ the perturbative beta function $\beta(g) = -[ b_0 g^3/(4\pi)^2 + b_1 g^5/(4\pi)^4 + \cdots]$  possesses, to 2-loop order, a non-trivial zero $g^*= -(4\pi)^2 b_0/b_1$ 
for a range of number of flavors $N_f^{**} < N_f < N^*_f$.  
%defining a ``conformal window" (CW). 
The upper end of this range, at which $b_0$ reverses sign,  is given by $N^*_f= {11\over 4\kappa} {C_2(G)\over T(R_f)}$ where    
$\kappa=1(1/2)$ for 4 (2 ) - component fermions. For $G=SU(N_c)$ with fundamental representation  fermions this is the well-known result $N^*=11N_c/2\kappa$. 
If $(N^*_f - N_f) << 1$ this zero is within the perturbative validity regime, and its existence can be trusted (Banks-Zaks IR FT) \cite{BZ}. The perturbative value of the lower end  $N_f^{**}$ , however,  given by the point where $b_1$ first reverses sign, cannot by trusted.  
Determining its actual (non-perturbative) value, i.e. the true extent of the ``conformal window'' (CW), 
is a question that has been intensively investigated in recent years for a variety of fermion representations $R_f$ and mostly $G=SU(3)$ or $SU(2)$ \cite{DelD1}. 
In this connection, it has been commonly assumed, e.g. \cite{MY}, that chiral symmetry will eventually be broken regardless of the number of fermion flavors provided the coupling is taken strong enough. In fact, it was recently found that this is incorrect.  
MC simulations for $N_c=3$ at vanishing or small inverse gauge coupling $\beta$  showed that chiral symmetry is restored via a first-order transition above a critical number of flavors ($\sim 52$ in the continuum) \cite{dFKU}. The same result was arrived at by resummation of the hopping expansion in the strong coupling limit: the familiar chiral symmetry breaking solution abruptly disappears above a critical $N_f/N_c$ \cite{T1}.  
Putting the available information together suggests a gross picture of the phase diagram of $N_f$ versus bare coupling $g$ at fixed $N_c$ shown in Fig.\ref{Fig1} below; the case of  
fundamental representation $SU(N_c)$ fermions is chosen for definiteness, other cases being qualitatively similar.

As just noted, the boundary separating the chirally broken phase from the chirally symmetric phase terminates at infinite coupling at a finite critical $N_f$ \cite{dFKU}, \cite{T1}. At the other extreme at vanishing coupling it terminates at a critical $N_f^{**}$ ($\sim 12$ for $N_c=3$) whose exact value remains somewhat controversial \cite{DelD1}. 
The boundary is known to be a first order phase transition at least for some range starting from the strong coupling limit end ($g\to \infty$). Also,  several studies over the years \cite{Misc} at fixed $N_f$ inside the CW have observed 
a first order phase transition to a chirally broken phase as the coupling is increased,\footnote{Additional, likely spurious, transitions may appear though at intermediate coupling depending on the particular fermion lattice action being used.} which is consistent with the picture in Fig. \ref{Fig1}. 
The simplest scenario would then be 
that the transition line is first order everywhere, though this might in fact depend on the 
fermion representation and gauge group. (Incidentally, a first order transition would be problematic for standard walking TC.)  

The region enclosed by the boundary between the chirally broken and unbroken phases and the horizontal broken line in Fig. \ref{Fig1} is then the putative CW. 
The standard picture of the RG flows  in the plane of irrelevant couplings inside the CW is shown in Fig. \ref{Fig2}-left. 
\begin{figure}[ht]
\begin{center}
\includegraphics[width=8cm]{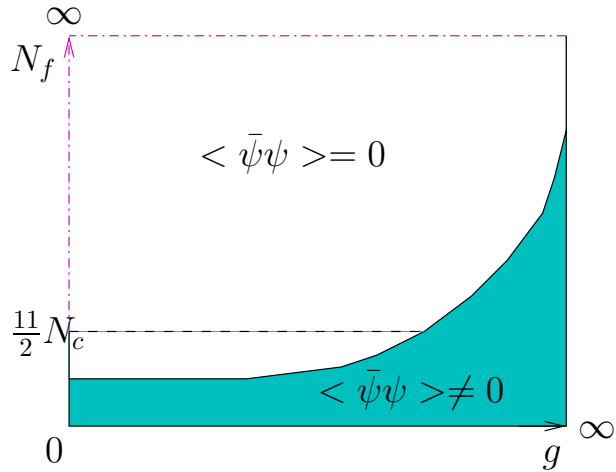}
%\hfill
 \end{center}
\caption{Phase diagram of $N_f$ vs. bare coupling $g$.
\label{Fig1}
 }
\end{figure}

As $N_f\to N_f^*$ from below the non-trivial IR FT $g_{\rm IR}^*$ inside the CW moves and eventually merges with the UV FT at the origin. For any $N_f > N_f^*$  
perturbation theory (PT)  gives a trivial IR fixed point at $g=0$. 
The simplest scenario  then would be  that no other FT is encountered at non-zero coupling (Fig. \ref{Fig2}-right).

Life in the region $N_f > N_f^*$, however, could turn out to be more exciting. Simulation studies measuring toleron mass, Dirac spectrum and hadron spectrum for $N_c=3$ at zero or small $\beta$ performed in \cite{dFKU} find evidence for a nontrivial IR FP in the region  $N_f > N_f^*$. Based on these measurements 
the conjecture was made in \cite{dFKU} that the FP location varies continuously with $\beta$, as well as $N_f$, reaching the value zero for $\beta\to \infty$, $N_f > N_f^*$, and for $N_f\to \infty$.  This would amount to a line of IR fixed points as depicted in Fig. \ref{Fig3}-left. 
Such a line, however, would seem to contradict weak coupling PT where no FP line ending at the FT at $g=0$ is seen. 
If such non-trivial IR FP's actually exist, their existence can be reconciled with PT if an even  
zero of the beta function obtains as depicted in Fig. \ref{Fig3}-right. Actually, one would expect such a zero to be unstable under changes in $N_f$ or other parameters unless perhaps it is an infinite order zero. 
More generally, though, this zero could appear as the limiting case of the situation shown in Fig. \ref{Fig4}-left. Here the possibility of other relevant directions (in addition to mass) is considered. 
These can arise from operators, such as chirally symmetric  4-fermi interactions, e.g., $G(\bar{\psi}\gamma_\mu\psi)^2$, whose anomalous dimensions at some intermediate couplings are such that they become relevant (marginal).  Massless quenched $\rm {QED}_4$ provides an example \cite{LLB}. 
As $N_f$ or other parameters are varied the non-trivial IR FT eventually merges with the non-trivial UV FT leading to the even degree zero in Fig. \ref{Fig3}-right.  Upon further increase of $N_f$ 
this zero disappears resulting in the situation in  Fig. \ref{Fig2}-right, and consistent with the fact \cite{dFKU}, \cite{T1} that for $N_f\to \infty$ at fixed $N_c$ the theory becomes trivial. The situation depicted in Fig. \ref{Fig4}-left provides one possible scenario for reconciling the FT's found in \cite{dFKU}, coming from the strong coupling side, with weak coupling PT. It is, however, not the only one (cf. below). 
\begin{figure}[t]
%\begin{center}
\includegraphics[width=7cm]{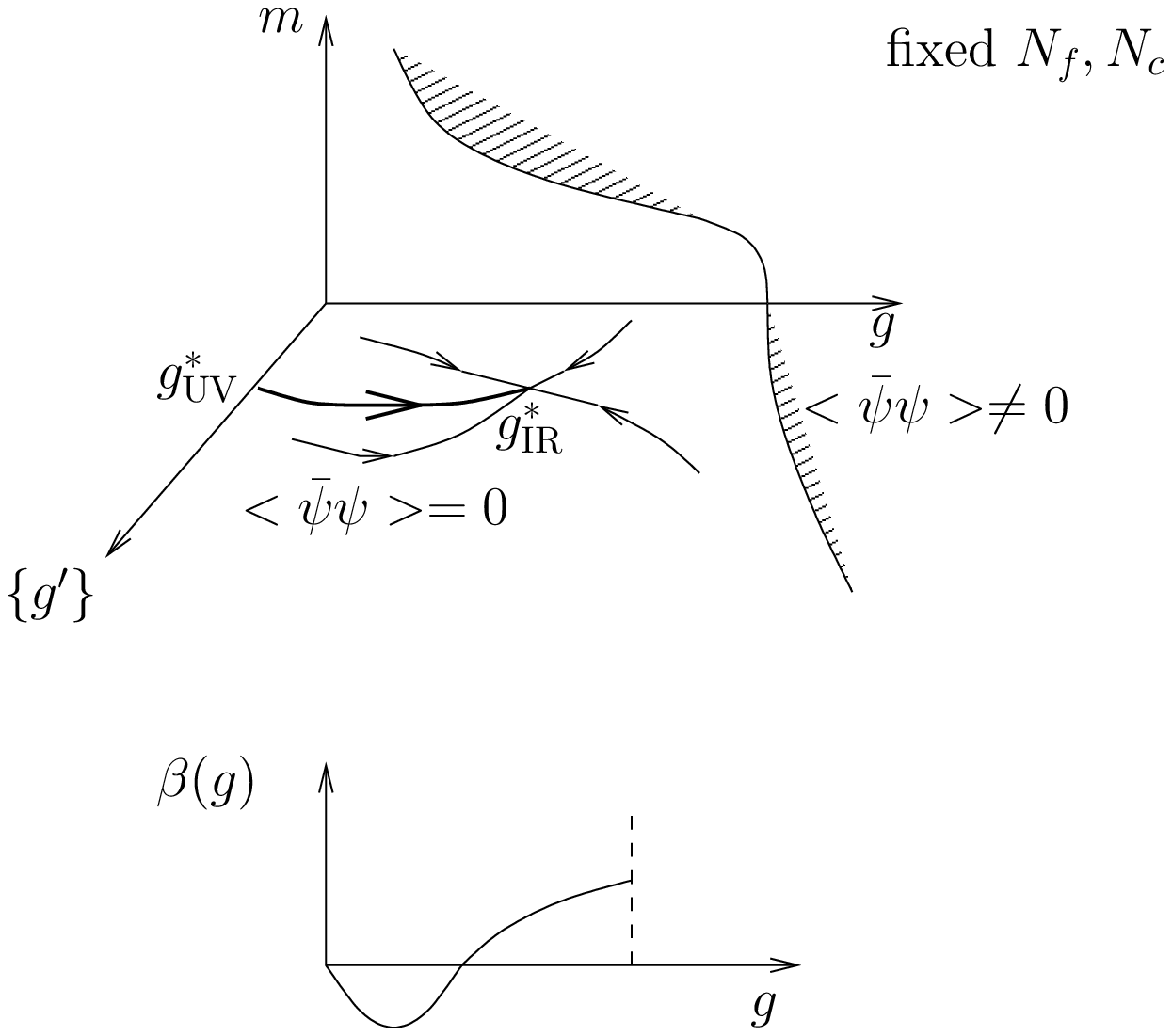} \hfill
\includegraphics[width=6cm]{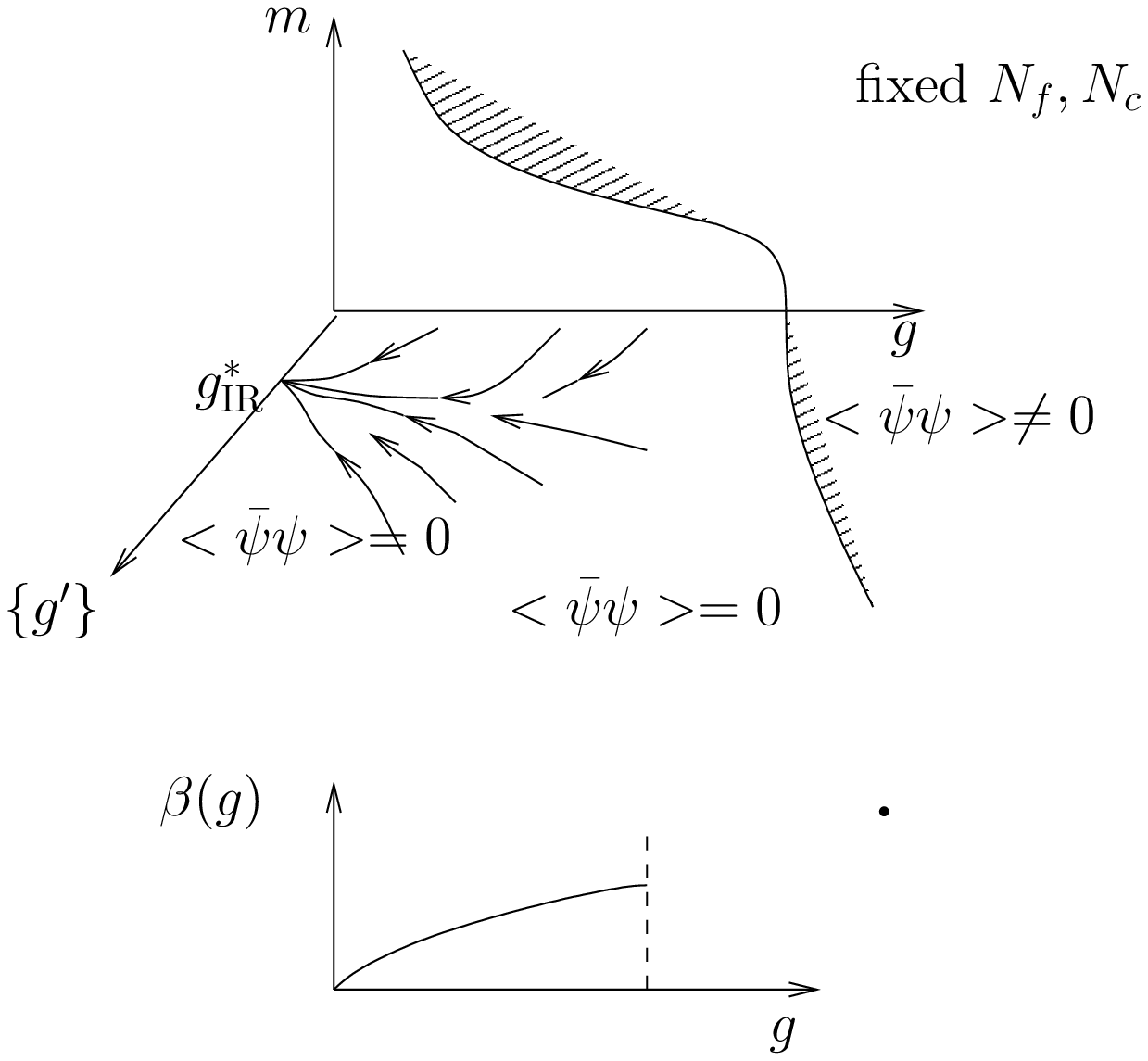}
%\includegraphics[width=6cm]{lat13fig8.eps}
%\end{center}
\caption{ Left: Standard picture within the putative CW ($N_f <  N_F^*$). $\{g^\prime\}$ denotes the set of irrelevant couplings and mass $m$ is the relevant direction. Right:
Lone trivial IR FT above the CW ($N_f > N_F^*$). }\label{Fig2}
\end{figure}

\begin{figure}[ht]
%\begin{center}
\includegraphics[width=6cm]{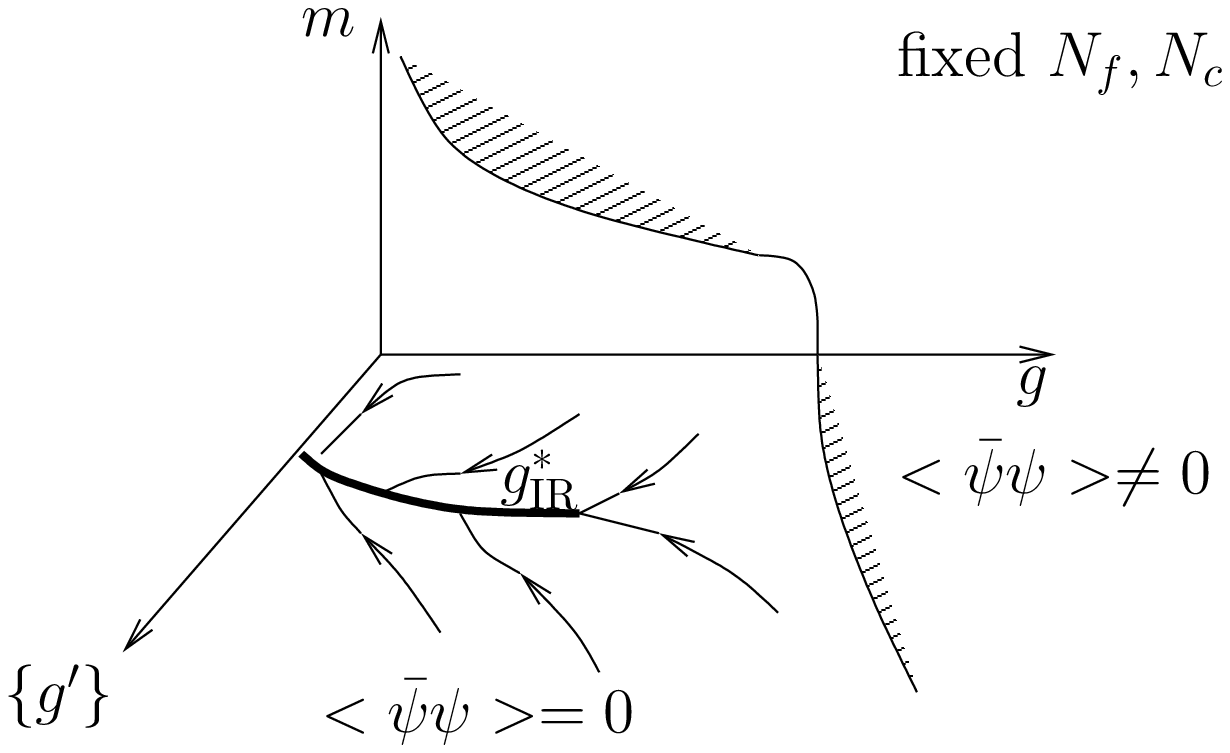}\hfill
\includegraphics[width=6cm]{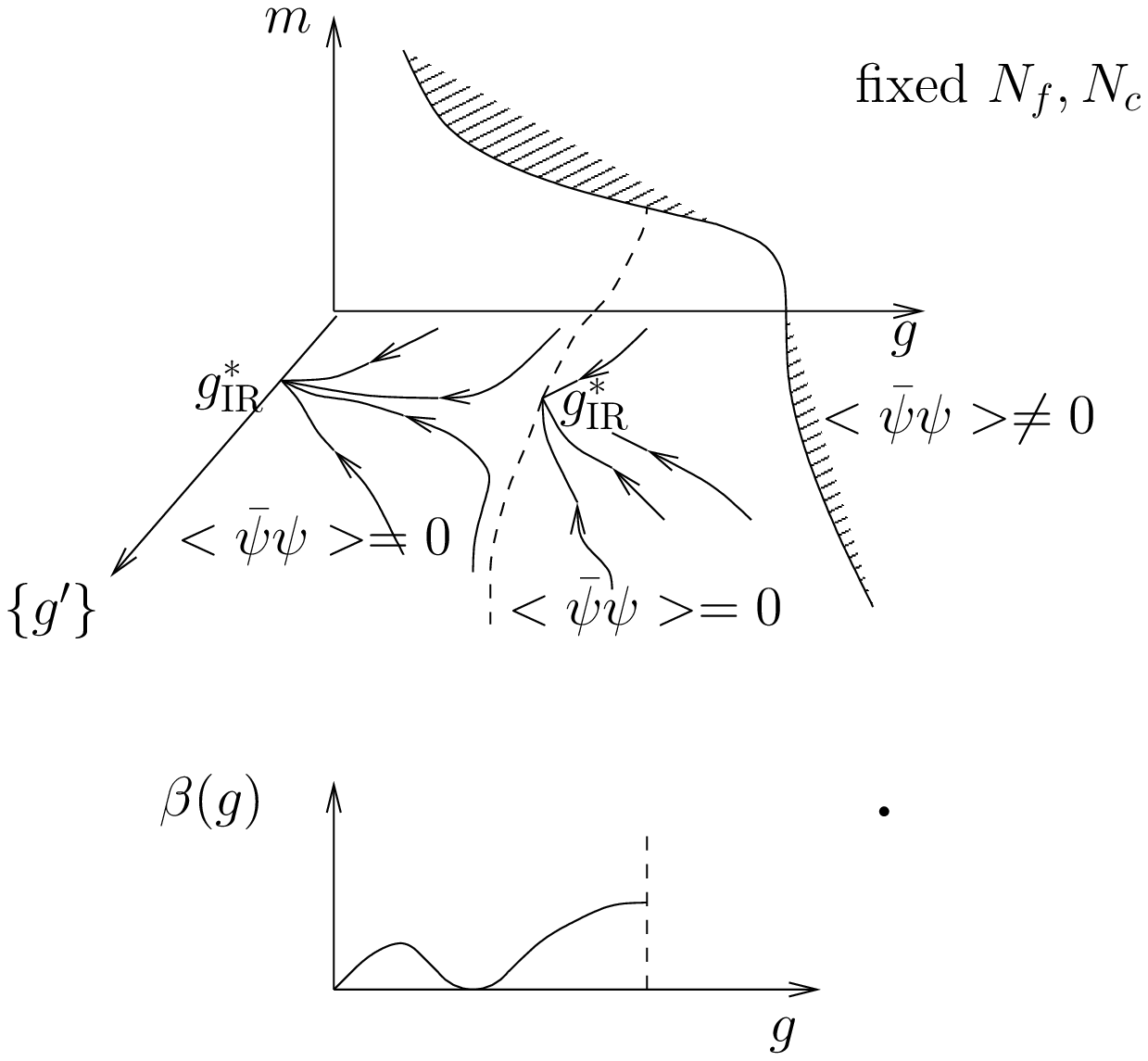}
%\includegraphics[width=6cm]{lat13fig9.eps}
%\includegraphics[width=6cm]{lat13fig7a.eps}
%\end{center}
\caption{Here $N_f > N_F^*$. Left: Line of IR fixed points with trivial end-point. 
Right: Non-trivial IR FT corresponding to a beta-function even zero. } \label{Fig3}
\end{figure}

\begin{figure}[ht]
%\begin{center}
%\includegraphics[width=6cm]{lat13fig2.eps}\hfill
%\includegraphics[width=6cm]{lat13fig3.eps}
\includegraphics[width=6cm]{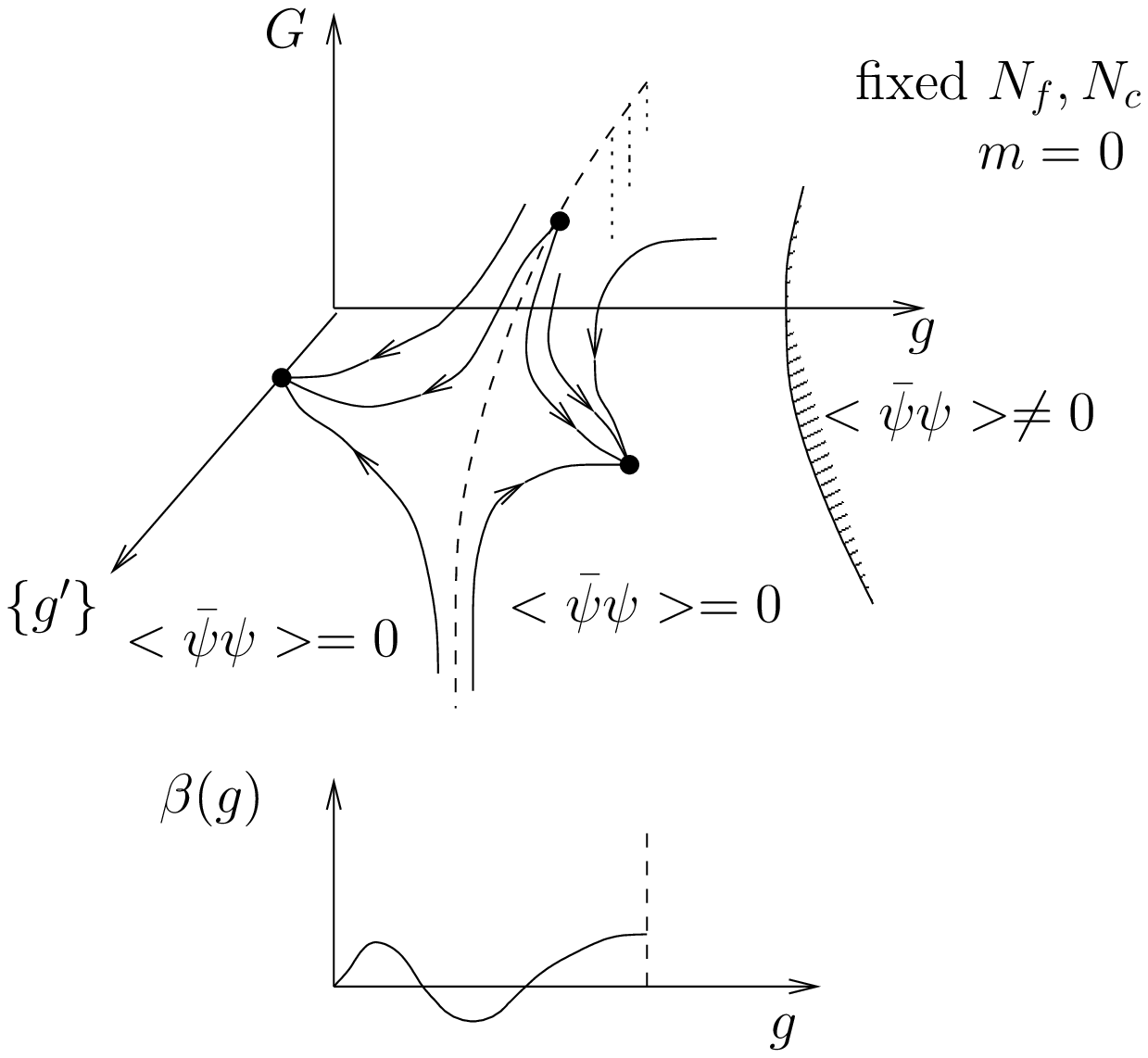}\hfill
\includegraphics[width=6cm, height=5.5cm]{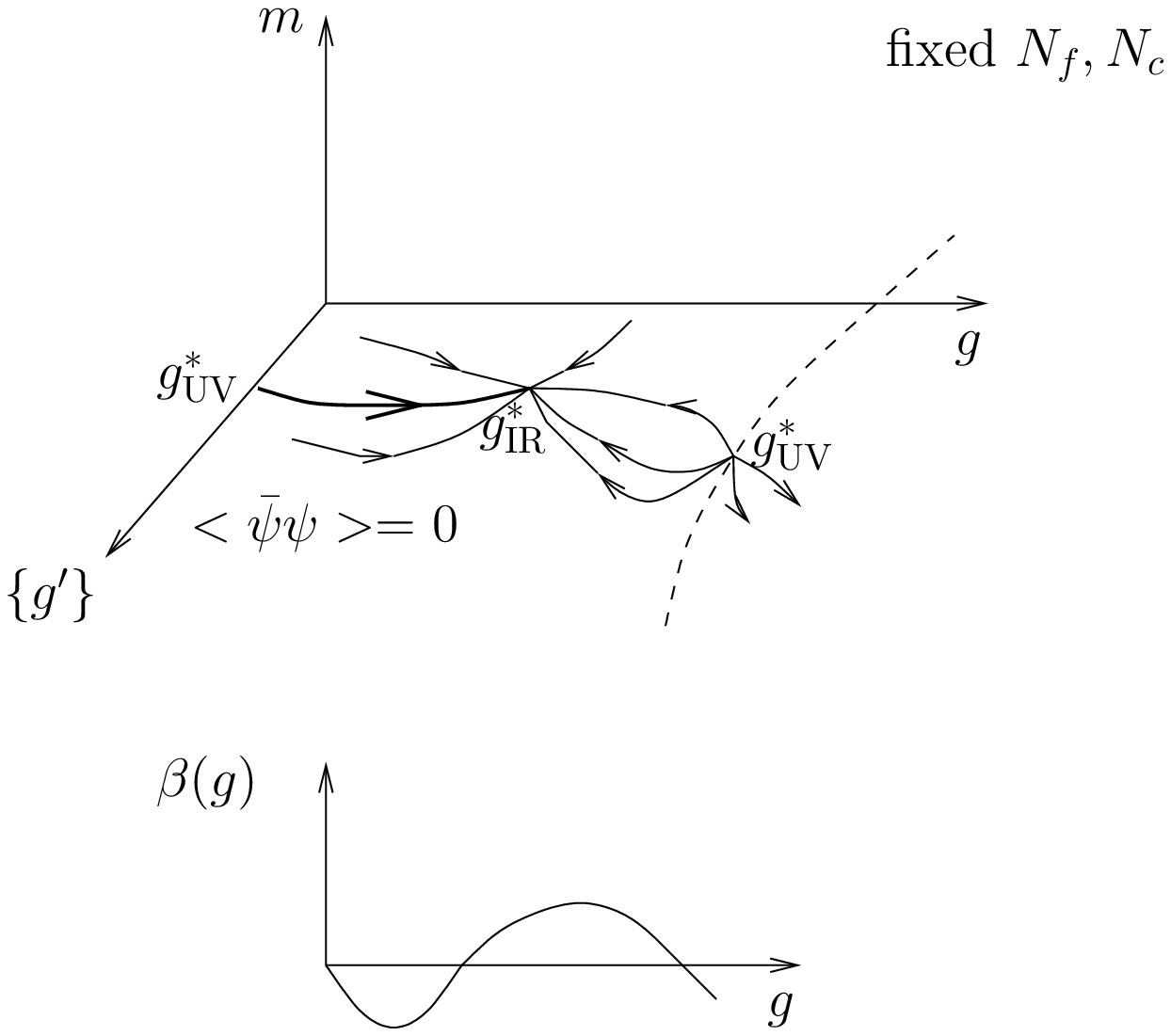} 
%\includegraphics[width=6cm]{lat13fig7a.eps}
%\end{center}
\caption{Left: Additional relevant direction $G$ resulting in nontrivial UV and IR fixed points for $N_f > N_F^*$.} Right: Analogous situation with a  non-trivial UV FT for $N_f < N_F^*$. \label{Fig4}
\end{figure}

The analog of the FP structure in Fig. \ref{Fig4}-left could actually arise also {\it within} the CW but, of course, with the sign of the beta-function reversed. This was discussed in \cite{KLSS}. It would amount to the occurrence of an UV FT beyond the non-trivial IR FT of Fig. \ref{Fig2}-left, as depicted in Fig. \ref{Fig4}-right. 

It should be emphasized in this connection that existing lattice simulations exploring IR conformal behavior are all at fixed low $N_c$, typically 2,3.  They do not explore the regime of both $N_f, N_c$ large, 
as, for example, when they are adjusted so that a BZ-like IR FT occurs at large $N_f$. This is precisely one of the regimes of interest for some of these new possibilities. Some indications already appear in considering the zeros of the beta-function 
in PT.  A general exploration of the non-trivial IR and UV zeros of the 4-loop beta function for a variety  of theories and representations is given in \cite{PS}. As always the question is whether  
such zeros can persist in the full theory. Typically, given a  zero found at a fixed order, the  neglected subsequent order terms in the beta function expansion, when evaluated at the location of the zero, can be as large as the retained terms.  This can happen even when this zero location goes as $\sim N_f^{-1}$ for large $N_f$.
Thus the unknown higher order corrections cannot be a priory neglected, and so, in contrast to the BZ FT, the existence of such zeros, even though perhaps suggestive, is not assured.

An alternative approach to computation of the beta-function in the continuum theory going beyond standard weak coupling PT is by means of an $1/N_f$ expansion at fixed 'tHooft coupling $ N_fg^2$ \cite{H}. 
The expansion amounts to a loop expansion with a modified gauge boson propagator dressed by the fermion bubble contribution to the self-energy, and modified $n$-point gauge boson vertices which include the contribution of the fermion loop with $n$ external bosons. 
%; and all gauge boson interaction legs carrying $1/\sqrt{N_f}$ factors. 
The expansion of the beta function has the form 
\beq 
\beta(\lambda) = {8\over 3}\lambda^2 \left[1 + \sum_{k=1}^\infty {H_k(\lambda)\over N_f^{\;k}} \right]    \, , \label{lNbeta} 
\eeq 
where $\lambda \equiv  \kappa T(R_f) N_fg^2/(4\pi)^2$. 
Each $H_k(\lambda)$ represents the contribution of a class of graphs, consisting of all graphs of the same, fixed $N_f$ dependence but including all orders in $\lambda$. 
Because of the non-local nature of vertices and inverse propagators in this expansion such computations generally cannot be performed exactly beyond leading order. For $H_1$, however, one has $H_1(\lambda) = - 11C_2(G)/(3\kappa T(R_f)) + \tilde{H}_1(\lambda)$, where, remarkably, an exact integral representation can be given for $\tilde{H}_1(\lambda)$ (for $SU(N_c)$ and also $U(1)$) so that its singularity structure can be deduced. 
Furthermore, the first several terms in the expansion of $H_k(\lambda)$, for $k\leq 4$, in powers of $\lambda$ are known \cite{H}.  If all $|H_k(\lambda)|$ are bounded at a given fixed $\lambda$ convergence follows for sufficiently large $N_f$. The $H_k$, however, appear to generally possess  (pole or log) singularities as functions of $\lambda$. On one side of a pole singularity (or either side of a log singularity) of some $H_k$ the beta function (\ref{lNbeta}) possesses a IR or UV zero (depending on the pole residue sign).  $\beta(\lambda)$ thus exhibits distinct branches, each branch delineated by a pair of singularities of the set of the $H_k$'s in a manner completely analogous to that found in the case of the supersymmetric pure $SU(N_c)$ beta-function \cite{KS}.  Within each such branch the beta function is given, for sufficiently large $N_f$, by a convergent series (\ref{lNbeta}). 
For $H_1(\lambda)$ the singularity structure can be deduced from the available exact integral representation, but for higher $H_k$'s only partial information is available. \cite{H} gives a summary of what is known or can be reasonably conjectured, see also  \cite{PS}, \cite{S}. For many purposes it would be sufficient  to just know the location of the first, or first two (i.e., lowest in $\lambda$) singularities in (\ref{lNbeta}).  In particular, for $SU(3)$ theories, $H_2$ is surmised to possess a pole singularity lower than the 
$H_1$ log singularity, and resulting in an IR FT just above it. This would be exactly the structure 
needed to provide an explanation of the IR FT's found in \cite{dFKU} coming from the strong coupling side. Again, however, it is hard to assess the reliability of these finding for the exact theory as the location and nature of the singularities delineating branches can be 
drastically altered by higher omitted contributions.  
The density of poles from the complete set of the $H_k$'s is not known, and the possibility also exists that a series of poles might sum up to an essential singularity.

At this point it is worth remarking that all cases investigated so far form only a small subset of possible IR-conformal gauge theories. In particular, only simple color groups have been considered. 
If the color group is semi-simple, e.g.,  $SU(N_1)\times SU(N_2)\times \cdots \times SU(N_n)$,  
there are $n$ gauge couplings resulting in a coupled set of equations for their beta functions.  
There are now correspondingly many choices for the coupling of fermions. Different fermion subsets may be coupled to different subsets of group factors and in different representations.  Depending on the number of such parameters available, many more possibilities for non-trivial IR and UV FT's may now arise. Investigations of FT's of such coupled beta function equations can easily become quite involved and, so far, have not been carried out in interesting cases even within weak coupling. 
In particular, existence of any non-trivial FT's at (relatively) weak couplings in such more general theories could turn out to be important for electroweak phenomenology.  

\section{Spectrum in IR-conformal theories}
Lattice simulations in IR-conformal or near conformal theories, and spectrum computations in particular, are very challenging. 
Asymptotic scale invariance is explicitly broken by the finite lattice size and non-vanishing mass for the fermions at which the simulations are necessarily carried out, as well as, in the case of Wilson fermions, by the non-chiral discretization of the Dirac operator. If the putative IR FP has 
additional relevant directions the finite lattice spacing may be another source of explicit deformation. 
Extrapolation to the chiral/conformal limits from a region of sufficiently small fermion masses and large enough lattices must be guided by some analytical picture: chiral PT in the case of QCD-like theories, and onset of some hyper-scaling regime for IR-conformal theories. Distinguishing between the two cases convincingly can be very tricky. 
Keeping systematic errors and in particular finite size effects under control, especially in the computation of gluonic spectra, can be very expensive. Progress, however, has been achieved in recent years in state-of-the-art large-scale computations  allowing some picture of the spectrum in IR-conformal theories to emerge \cite{Sp1}, \cite{Sp2}, \cite{Sp3}.  

Relevant parameters away from conformality are a 
quark mass $\hat{m}=m/\mu=am$, and the lattice size $L$.  
(Coupling to other, extraneous gauge interactions may of course provide other explicit deformations.)  
The basic picture one has of IR-conformal behavior is as follows. There is a "locking" scale $M_l$ (which depends on the specific theory dynamics) below which a scaling regime obtains where physical mass ratios remain essentially constant. 
Hadron masses scale as \quad $M_{\rm H} \sim \mu \hat{m}^{1/(1+ \gamma_m^*)}$ \cite{DelDZ}, with mass anomalous dimension $\gamma^*_m$ at the FP. 
The actual ordering of the spectrum, which is theory-dependent in its exact detail, is essentially set at this scale. If $M_l$ is relatively high, one expects a spectrum ordering similar to that of QCD with heavy fermions, i.e. its lower part consisting of a nearly degenerate meson spectrum above a still lower set of glueball states with $0^{++}$ being the lowest. If $M_l$ is low, one may expect a spectrum where the pseudoscalars are lower than the vector mesons and roughly at the same level as the low gluonic states. 
As $m$ is lowered below the locking scale $M_l$ then, the spectrum (including the string tension which is always the lowest level) scales toward zero according to the above relation. 
Simulations in the case of $SU(2)$ with two flavors of adjoint Wilson fermions \cite{Sp2}, and for 
$SU(3)$ with 12 flavors (three degenerate staggered fermions) \cite{Sp3} favor the first scenario. 
The lowest states are the gluonic states, with $0^{++}$ being the lowest, and the meson states clearly  separated above. In the $SU(3)$ case the flavor singlet scalar $0^{++}$ meson is lowest and found to be very close to the gluonic $0^{++}$ state, these two then forming the lowest states. (Flavors singlets where not computed in the $SU(2)$ case.) 
This picture was in fact first suggested from analytic considerations around a BZ FT in \cite{M1}, and, at least in the cases considered, according to these lattice simulations appears to hold also for a non-perturbative IR FP. 

In a phase governed by an IR FP, as all relevant deformation parameters are switched off 
($m\to 0 $, $L\to \infty$),  the spectrum collapses to only massless states (``unparticles"). 
But at any small non-zero deformation such as a non-vanishing $m$ or finite box size one has a particle spectrum, with mass gap as sketched 
above, and, according to the above findings, containing a light scalar $0^{++}$ meson and a scalar $0^{++}$ glueball state plus the (somewhat heavier) rest of the  glueball and meson/baryon spectrum. 
%\[ \sim \bar{\psi}\psi, \quad \bar{\psi}\gamma_5\psi, \quad \bar{\psi}\gamma_k\psi, 
%\quad \cdots  \]

The presence of naturally light scalar composite states in the spectrum of IR-conformal theories provides a natural setting for consideration of composite Higgs models.

\section{Composite Higgs in IR conformal theories} 

Consider a  theory with $N_f$ `techniflavors' and  $N_c$ `technicolors' such that its IR behavior is controlled by an IR FP.   This IR FT may arise from any of the situations reviewed above, either inside the so-called CW or above it. For phenomenological reasons we would generally prefer it to be at weak coupling. Note, however, that, as seen from our previous discussion, this does not necessarily mean that the formation of the composites in the spectrum originated in a weak coupling regime. 
As we also saw, this may imply that $N_f$ and $N_c$, must be suitably adjusted and be large. 
It is expedient to consider a non-simple color group of at least two factors as this most naturally can accommodate a dark sector - see below. The basic idea we want to suggest here is the following. 

In the presence of a small relevant deformation (e.g., a small quark mass $m$ or large finite box size $L$) one has a well-defined discrete spectrum of composite states. We make the assumption that, in accordance with the available computations, the scalar states are the lightest states in the spectrum. Coupling next other gauge interactions, 
in particular electroweak interactions, renders this system of (arbitrarily) light scalar states unstable under the Coleman-Weinberg  mechanism. The resulting mass gap is now in effect the dynamically generated 
conformality deformation, which persists in the limit where the original explicit deformation is removed  ($m\to 0 $ or $L\to \infty$).  
This coupling of electroweak interactions directly to the naturally light (massless) states present in an IR-conformal theory allows for a wide class of potential composite Higgs models.

To sketch an example of this type of model, consider an IR-conformal theory 
%$SU(N)$ (or $U(N)$) 
with $N_f$ flavors in the fundamental representation of the color gauge group.  
%such that system in CS phase. 
Single out just two of these flavors $Q=(U,D)$.  
%(to be later coupled to the electroweak $SU(2)\times U(1)$). 
There are now composite  states formed by $Q$ and the remaining fermion flavors $\psi_a$, $a=1, \ldots, 
N_f-2$, such as $ \bar{\psi}\psi$, $\bar{\psi}Q$, $\bar{Q}Q,  \ldots$. 
The ``mixed" sector can be eliminated by taking semi-simple color gauge group, 
e.g., $SU(N_1)\times SU(N_2)$ with $\psi$ charged under both factors, and the $Q$ charged under only one factor. 
 `Mixed' composites such as $\bar{\psi}Q$, $\psi QQ, \ ...$ no longer form. The only possible color singlet mixed states that could form are highly unstable multi-quark (tetra and higher) states if they form at all.  
 
 Now consider coupling the electroweak interactions. This may of course be done in various ways depending how they are to be aligned relative to our IR-conformal theory. Here we just couple to the singled-out $Q$ fermions as follows.  
The scalar meson $\bar{Q}Q$ gives rise to the four fields: \\ 
%$h_i, i=0, 1, 2,3$: 
$ h^+ = - \bar{D}U,  \quad h^-= \bar{U}D, \quad h_3= (\bar{U}U - \bar{D}D)/\sqrt{2}, 
\quad  h_0= (\bar{U}U + \bar{D}D)/\sqrt{2}  $.  
These may be taken to form the weak scalar doublet 
\beq \quad H= \left(\begin{array}{c} h^+ \\ (h_0+ ih_3)/\sqrt{2}\end{array} \right) \; , \qquad 
\tilde{H}= i\tau_2 H^*= \left(\begin{array}{c} (h_0- ih_3)/\sqrt{2} \\ h^- \end{array}\right) 
\eeq  
after giving $Q$ ordinary quark elw charges. 
In addition one has, of course, the other pseudoscalar $P= \bar{Q}\gamma_5 Q$, 
vector $V_k= \bar{Q}\gamma_k Q$, etc., meson states, as well as baryon states. 
The glueball states are all weak singlets. In particular one has the $0^{++}$ 
glueball state, which,  together with the $h_0$, are expected to be the lightest states. These two scalar states may in general mix. If the mixing is small, the predominantly fermionic 
composite can serve as the physical Higgs, whereas the predominantly gluonic component can serve as a nearly `invisible' weakly interacting particle of comparable mass (WIMP). Another phenomenologically interesting scenario is the case of the fermionic scalar composite being not too different in mass from 
the pseudoscalars and other mesons, while  the gluonic scalar mass is rather lower (more like the $SU(2)$ spectrum case). Mixing then can result into a lower mass scalar (Higgs) and a second heavier physical scalar among the other massive meson states. Such detailed dynamical questions as the exact mass splittings among the light states and the amount of mixing are specific theory dependent and can only be answered by actual computation. 

In addition one has of course the  `dark' sector containing the scalar $\Psi=\bar{\psi}\psi$ and the other meson and baryon states formed among the remaining flavors $\psi_a$ which carry  
no electroweak charges. 

It would be interesting to consider replacing one or both 
factors in $SU(N_1)\times SU(N_2)$ by $U(N_1)$, $U(N_2)$ factors which would eliminate the baryon states in one or both sectors. For sufficiently large number of colors and flavors this 
would be expected to otherwise make little qualitative difference, but again there are no available simulations of IR-conformal theories with $U(N)$ color groups.

The effective theory of the IR-conformal interactions at low energies is in principle obtained by matching the composites to interpolating fields. Schematically, this will result in an effective potential of the form 
\bea 
 \lambda (H^\dagger H)^2 + \lambda_1(P^\dagger P)^2 + \lambda_2 |P^\dagger H|^2 + \cdots 
\lambda_V |V^\dagger_kV_k|^2 
%& &  \nonumber \\ 
+ \cdots  + \lambda_d (\Psi^\dagger\Psi)^2  + 
  \lambda_d^\prime (\Psi^\dagger\Psi)(H^\dagger H) + \cdots  \label{effpot}
  %& & 
\eea  
If the  IR FP occurs at relatively weak coupling, all effective couplings in this long distance effective potential are correspondingly weak. Note, however, that, though desirable for computational or other reasons, such weak couplings are not a fundamental requirement. Also note in this connection that all available simulations in various IR-(nearly) conformal models give small anomalous dimensions. 
This implies that their contributions through the interpolating field matchings lumped into the effective couplings $\lambda_i$ in (\ref{effpot}) are small.  
The coupling to the electroweak gauge fields then renders this system of (nearly) massless scalar fields 
unstable under the Coleman-Weinberg mechanism. 
With parity and Lorentz symmetry assumed preserved, as usual, only the scalar  $H$ condenses 
in the resulting breaking which will give the usual electroweak breaking pattern. From the available deformed spectra computations it is not hard to envision that a gap of a factor of, say, 5-10 can develop between the lightest massive scalar (physical Higgs), plus possibly a WIMP-like particle, and the higher massive states resulting from this breaking. 
Interactions with the dark sector enter naturally here through the `Higgs portal'. 

For the dark sector itself one may introduce other `dark' gauge interactions among the $\psi_a$'s so that, through Coleman-Weinberg, fermion condensates or other mechanisms, it is rendered massive or consists of massive and unparticle sectors.  

Masses for the ordinary SM quark could be introduced by the usual mechanism of effective 4-fermi interactions  between the $Q$ fermions and the SM quarks  at a high scale. The interpolating field matching to the $Q$-composites converts such interactions to effective Yukawa couplings for the SM quarks. These provide the quark masses as in the SM after the composite $H$ field condenses via the Coleman-Weinberg mechanism. (Note that no techniquark condensate, as in extended TC, at some intermediate scale, with its attending fine-tuning problems, need here be invoked.) 
The number of boson degrees of freedom present generally suffices to allow the Coleman-Weinberg mechanism to be implemented also after the introduction of the 
quarks. The effective theory describing such models would in fact be similar to  many models in the literature with elementary scalar fields and 
electroweak breaking \`{a} la Coleman-Weinberg in the presence of a dark sector driven by potentials of the type (\ref{effpot}) \cite{HT}.

\section{Conclusions}

The search for non-trivial FT's in gauge theories is still at an early stage. Despite considerable efforts already devoted to it, there are, as we saw, many unanswered questions concerning the occurrence 
of IR, but also UV, nontrivial FT's both inside and outside the putative "conformal window" in 
non-abelian theories with simple color group. More general gauge theories coupled  
to large number of fermion flavors have in fact hardly been considered. 
In few cases where the existence of an IR FT can be reasonably ascertained, however, lattice computations of the spectrum have provided significant findings, in particular, the non-QCD-like mass ordering of 
states with the scalar states being the lightest. As suggested here, these naturally light composite scalar states (massless in the limit of removing any relevant conformal deformation) provide the setting for 
models of electroweak breaking where the Higgs is one (or a linear combination) of these 
composite scalars. These then are composite Higgs models very different from the Higgs as NG boson or walking TC models. As noted above their effective description would be more similar to that of models with scalar (but now composite as opposed to elementary) fields and Coleman-Weinberg 
breaking in the presence of a dark sector. 

In summary, it should be clear that existence of IR-conformal theories and the associated presence of light composite scalars opens up the possibility for a variety of new composite Higgs models. The real difficulty is identifying the FT structure and spectrum of a candidate IR-conformal theory and extracting the quantitative information from it needed for model building in each particular instance. 
With presently available techniques, this generally requires extensive computational 
effort.    

\vspace{1cm}

The author would like to acknowledge the Aspen Center for Physics for hospitality and thank the participants of the of the workshop "LGT at the LHC era", in particular, F. Sannino, Y. Meurice, A. Hasenfratz, M. P. Lombarto, B. Lucini, E. Pallante,  G. Fleming and E. Neil for discussions.

\end{document}